\documentclass[12pt]{article}

\usepackage{times,color}
\usepackage{amsmath}
\usepackage{subfigure,epsfig,graphicx,epstopdf}
\usepackage[colorlinks]{hyperref}
\hypersetup{linkcolor = {blue},	citecolor = {blue},	urlcolor = {blue}}
\usepackage{cite}
\usepackage{caption}

\newenvironment{sciabstract}{%
\begin{quote} \bf}
{\end{quote}}

\newcounter{lastnote}

\begin{document}

\title{A Test of Empty Wave via Quantum Memory in a Weak Measurement Scheme}
\author{Jian-Peng Dou$^{1,2}$, Feng Lu$^{1,2}$, Hao Tang$^{1,2}$, Xiao-Wen Shang$^{1,2}$, Xian-Min Jin$^{1,2,3,4\ast}$ \\
\normalsize{$^1$Center for Integrated Quantum Information Technologies (IQIT), School of Physics and Astronomy}\\
\normalsize{and State Key Laboratory of Advanced Optical Communication Systems and Networks,} \\
\normalsize{Shanghai Jiao Tong University, Shanghai 200240, China}\\
\normalsize{$^2$CAS Center for Excellence and Synergetic Innovation Center in Quantum Information}\\ 
\normalsize{and Quantum Physics, University of Science and Technology of China, Hefei, 230026, China}\\
\normalsize{$^3$TuringQ Co., Ltd., Shanghai 200240, China}\\
\normalsize{$^4$Chip Hub for Integrated Photonics Xplore (CHIPX),}\\ \normalsize{Shanghai Jiao Tong University, Wuxi 214000, China}\\
\normalsize{$^\ast$E-mail: xianmin.jin@sjtu.edu.cn}
}

\date{}

\baselineskip24pt

\maketitle

\begin{sciabstract}
In quantum mechanics, a long-standing question remains: How does a single photon traverse double slits? One intuitive picture suggests that the photon passes through only one slit, while its wavefunction splits into an ``empty" wave and a ``full" wave. However, the reality of this empty wave is yet to be verified. Here, we present a novel experimental configuration that combines quantum memory and weak measurement to investigate the nature of the empty wave. A single atomic excitation is probabilistically split between free space and a quantum memory, analogous to the two paths in a double-slit experiment. The quantum memory serves as a path detector, where single-photon Raman scattering is enhanced due to the presence of a stored spin wave, without collapsing the quantum state. This enhancement is recorded as classical information, and the spin wave stored in the quantum memory is retrieved twice, with an interference visibility of (79±2)\%. Unlike conventional weak measurement schemes, where weak values are detected during post-selection, our approach converts the weak value into classical information before interference takes place. Our results demonstrate the potential of quantum memory as a measurement device that preserves coherence while extracting partial information, offering new insights into quantum measurement.\\

\end{sciabstract}

\section*{1 Introduction}
Wavefunction is central to quantum theory \cite{Lundeen2011, Ringbauer2015}, yet its nature remains unclear. Many questions resolve around its physical meaning, particularly in explaining the ``which-way'' question in double-slit experiments. To determine the path of a quantum particle (such as a photon or an atom), a which-way detector is typically employed behind the slits, as suggested by Feynman \cite{Feynman}. However, acquiring path information destroys the interference pattern \cite{ Scully1991, Durr1998}. Currently, the notion of wavefunction collapse is commonly accepted: Before a measurement, the wavefunction is distributed over both paths the particle may take, and the quantum particle does not have a deterministic position; When the wavefunction is measured by a which-way detector on one possible path, it collapses instantaneously, projecting the particle onto one deterministic path. Consequently, there is nothing left for the surviving wave to interfere with effectively, and no interference pattern is observable. However, the physical process of instantaneous wavefunction collapse is still unclear and remains controversial. This issue, known as the measurement problem, represents a key tension in quantum physics \cite{Drossel2018, Bild2023}.

The above problem can be alternatively explained by the de Broglie-Bohmian (dBB) interpretation \cite{Bohm1952I, Oriols2019, Tumulka2021, Sanz2019, Wiseman1998, Kocsis2011, Braverman2013, Mahler2016, Xiao2019, Foo2022}. In this interpretation, a single quantum particle follows a deterministic trajectory and possesses a deterministic position at any given time, though practical measurements show statistical uncertainty due to the particle's initial position uncertainty. According to the dBB interpretation, in a double-slit experiment, a photon passes through only one slit per trial, but its wavefunction passes through two slits simultaneously, splitting into two waves: an empty wave (containing no photon) and a full wave (containing the photon). Neither of these two waves can be considered nonexistent, as their recombination produces an interference pattern. This phenomenon has been unequivocally demonstrated by numerous experiments, including single-photon interference in double-slit experiments, Mach-Zehnder interferometer experiments, a temporal Wheeler’s delayed-choice experiment \cite{Dong2020} and a twisted double-slit experiment with high-visibility HOM interference \cite{Zhou2017}.

Therefore, it is tempting to suspect that a separate empty wave might have an observable effect. Several experiments have been conducted to detect empty waves \cite{Wang1991, Croca1992, Zou1992, Jeffers1994, Folman1995,Muckenheim1988, Selleri1988, Broglie1968,Auletta04}, and some speculations about empty wave have been disproved, such as photon detection induced by an empty wave, attenuated empty waves, early arrival empty wave and coherence induced by an empty wave. As far as we know, the effect of empty wave is observable only when it is recombined with the full wave. However, after the recombination of empty wave and full wave, the empty wave is no longer empty, and the effect of a separate empty wave can not be directly demonstrated. To this day, the reality of the empty wave remains unclear, as does the instantaneous collapse of the wavefunction in the well-established conventional probabilistic interpretation.

\begin{figure*}
\centering 
\includegraphics[width=1\textwidth]{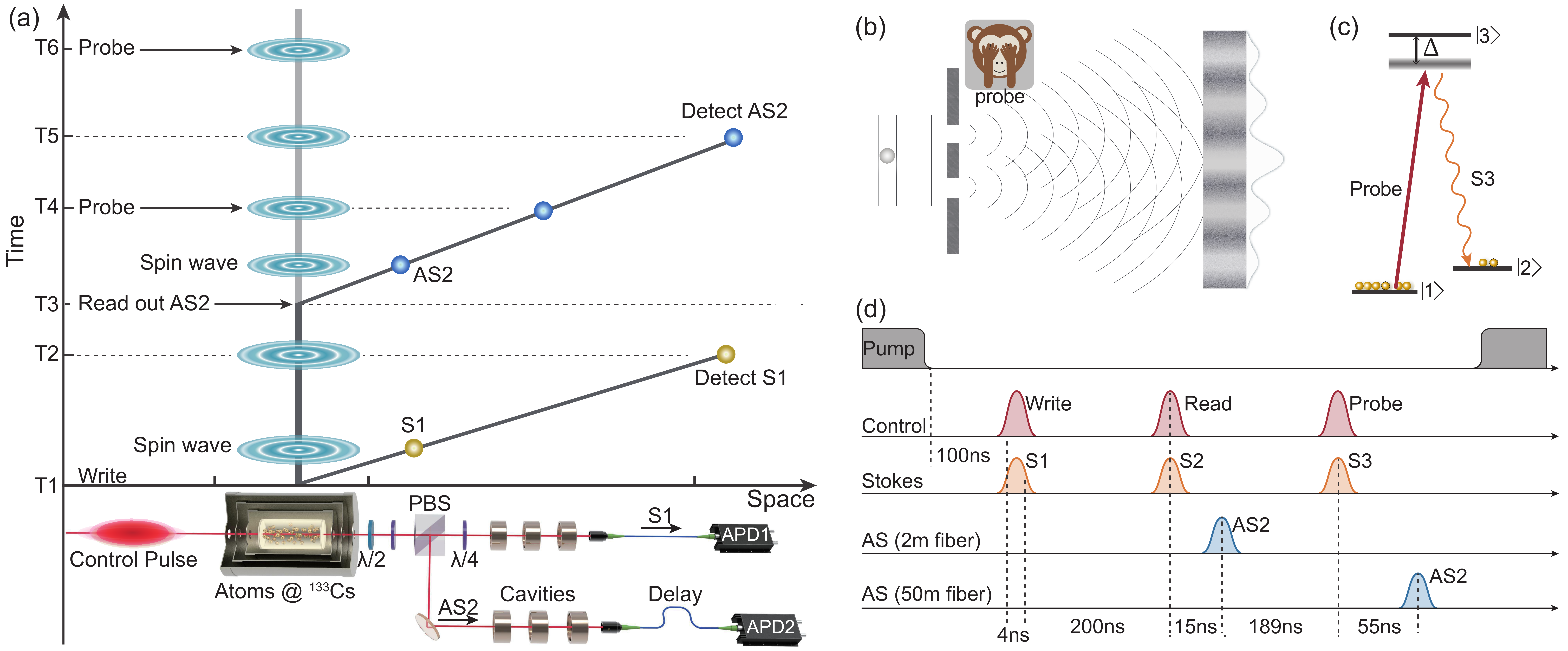}
\caption{\textbf{Schematic diagram of the experiment.} \textbf{(a)} Space-time diagram of the experiment. By splitting the spin wave into two parts—a spin wave stored in atoms and a propagating light wave—this light-matter hybrid interface enables the unconventional probing of empty spin waves. Stokes photons (S) are detected by APD1, and anti-Stokes photons (AS) are detected by APD2. $\lambda /4$: quarter wave plate; $\lambda /2$: half wave plate; PBS: polarization beam splitter; APD: avalanche photodiode. Before encountering the PBS, both S and AS are horizontally polarized by two wave plates $\lambda /2$ and $\lambda /4$. After passing through the PBS, both photons are circularly polarized by a 45° oriented quarter-wave plate. Then, S goes through the Stokes-resonant cavity, while AS is reflected by the front surface of the Stokes-resonant cavity, and then pass through the quarter wave plate once again. AS therefore flips its polarization to vertical polarization and is reflected by the PBS, and finally passes through the anti-Stokes-resonant cavity. See Appendix A and B for more details. \textbf{(b)} Conceptual diagram of the weak measurement. A peculiar experimenter (denoted by a monkey with his hands over his eyes) observes the double-slit experiment, which disturbs the wavefunction with a negligible effect in each trial. It cannot distinguish which slit the photon passes through due to he cannot clearly see the experiment. It may record some vague path information in a classical way, and some clues about the wavefunction may be recovered by a postselection on the wavefunction. \textbf{(c)} Illustration of the energy level for slightly probing the spin wave stored in atoms by Raman scattering. The detuning is set to ${\Delta=13.1\,{\rm GHz}}$. \textbf{(d)} Time sequence of pump light, control pulse, Stokes photon and anti-Stokes photon. The pump light prepares the atomic initial state $\left| {1} \right\rangle$. AS is delayed by a 2\,m fiber or a 50\,m fiber. Note that, when AS2 is delayed by a 2\,m fiber, the time delay between S2 and AS2 is 15\,ns rather than 10\,ns, which is due to that AS2 walk a longer distance in the air than the S2. }
\label{fig1}
\end{figure*}

Here, we experimentally test the reality of empty wave using a quantum memory enabled light-matter interface involving single photons and warm cesium atoms. The atoms not only serve as quantum memory capable of generating and storing the quantum wavefunction as an atomic spin wave, but also act as a beam splitter, dividing the spin wave into two parts: a spin wave stored in atoms and a propagating light wave. We utilize Raman scattering to slightly probe (i.e. weak measurement) the stored spin wave and demonstrate that our probe method does not collapse the wavefunction involving atoms and photons, unlike a conventional which-way detector.  We show that there is an anti-correlation between enhanced Raman scattering and an empty spin wave. Additionally, we demonstrate that our probe pulse bridges a superposition state to a strong measurement, and that it is the strong measurement which nonlocally projects the quantum wavefunction into a new state, rather than the probe pulse directly acting on the atoms.

\section*{2 Experimental results}

The schematic diagram shown in Fig.\,\hyperref[fig1]{1} illustrates the basic idea of our proposal and the configuration of the experiment. At time T1, the first control pulse named by write pulse enters the atomic ensemble, and generates a flying Stokes photon (S1) and a correlated collective atomic excitation in atoms simultaneously. At time T2, S1 is detected by a single-photon detector named by APD1, which heralds the existence of atomic spin wave. After a storage time of 200\,ns, at time T3, the second control pulse named by read pulse retrieves the stored atomic excitation as a flying anti-Stokes photon (AS2) with a probability around 10\%. If the atomic excitation is read out,  then an empty spin wave is stored in the atoms. At time T4, AS2 has not arrived at the detector APD2. At time T5, AS2 arrives at the detector and is detected. The third control pulse, named probe pulse, probes the effect of empty spin wave before or after the detection of AS2. The order of precedence between the probe pulse and the detection of AS2 is determined by the length of the fiber used to delay the detection of AS2.

We analogize the probe method to a monkey observing the double-slit experiment with his hands over his eyes in Fig.\,\hyperref[fig1]{1(b)}. Unlike a conventional which-way detector, our probe method records some vague information about the path information, and some clues about the wavefunction may be recovered by a postselection on the wavefunction. At the same time, the stored empty spin waves survives, facilitating our study of the reality of empty spin wave. The effective Hamiltonian of this light-matter hybrid interface is described by
\begin{equation}\label{hamiltonian}
H = \left( \lambda   \hat {S}^\dag   \hat {a}^\dag    +\lambda ^*  \hat {S}  \hat {a}   \right) + \left( \kappa \hat {S}^\dag    \hat {b}   + \kappa^* \hat {S}  \hat {b}^\dag    \right) ,
\end{equation}
where $\kappa$ and $\lambda$ are coupling parameters. $\hat {S}^\dag  $, $\hat {a}^\dag $ and $\hat {b}^\dag  $ are the creation operators for the atomic excitation, Stokes photon from Raman scattering and the retrieved anti-Stokes photon respectively. The first part of the above Hamiltonian corresponds to the write process or probe process, and is similar to a two-mode squeezing Hamiltonian \cite{Scully1997, Duan2001}. One can get a stored atomic excitation and a Stokes photon simultaneously by Raman scattering. In addition, a two-mode squeezing Hamiltonian implies an amplification for a preexisting photon or atomic excitation. By probing the amplification effect, one can nontrivially test the stored quantum state. Compared with an all-optical system, this is a unique advantage of a memory-enabled light-matter hybrid interface. The second part corresponds to the probabilistic retrieval in read process, and is known as a beam splitter Hamiltonian \cite{Hammerer2010}, implying that there is only one quantum: either an atomic excitation or a flying anti-Stokes photon.

As the read pulse energy is set to a modest value, the retrieval of AS2 is probabilistic. Similar to a double-slit experiment, the spin wave is splitted into two parts: a spin wave still stored in atoms and a flying light wave. The state involving atomic excitation and AS2 can be described by conditional functions \cite{Oriols2019}:
\begin{equation}\label{Bohm01}
\begin{aligned}
&c_1 \left| {\rm EW} \right\rangle _{\rm{Atom}}\left| 1 \right\rangle _{{\rm{AS2}}},  \\
&c_2 \left| 1 \right\rangle _{\rm{Atom}}  \left| {\rm EW} \right\rangle _{{\rm{AS2}}},
\end{aligned}
\end{equation}
where $\left| {\rm EW} \right\rangle _{\rm{Atom}}$ denotes the empty spin wave stored in atoms, $\left| 1 \right\rangle_{\rm{AS2}}$ represents the flying full wave that contains a flying anti-Stokes photon (AS2), $\left| {\rm 1} \right\rangle _{\rm{Atom}}$ denotes the full spin wave containing an atomic excitation in atoms, $\left| {\rm EW} \right\rangle _{\rm{AS2}}$ denotes a flying empty wave containing no anti-Stokes photon. $c_1$ and $c_2$ are proportional coefficients depend on the retrieval efficiency. 

Then, a probe process is applied to the residual spin wave still stored in atoms. Assume that an empty spin wave (no atomic excitation) can enhance Raman scattering, as well as enhance the creation of an atomic excitation, just as a full wave does. After the probe pulse, but before the detection of S3, the states in (\ref{Bohm01}) will change to:
\begin{equation}\label{Bohm02}
\begin{aligned}
&c_1\left| {2G - 2} \right\rangle _{\rm{S3}}\left| {2G - 2} \right\rangle _{\rm{Atom}}  \left| 1 \right\rangle _{{\rm{AS2}}}, \\
&c_2 \left| {2G - 2} \right\rangle _{\rm{S3}}\left| 2G-1 \right\rangle _{\rm{Atom}}  \left| \rm EW \right\rangle _{{\rm{AS2}}} ,
\end{aligned}
\end{equation}
where $G=\cosh ^2 (\kappa \delta t)$, and $\kappa$ is a coupling coefficient \cite{Louisell1961,Chen2009}, $\delta t$ is the width of probe pulse, see Appendix C for more details. The average number of S3 due to spontaneous Raman scattering in each probing trial is $\left( {G - 1} \right)$. Here, we use the average number $2G-2$ to denote the state of S3 after a trial of probe, which is similar for $\left| {2G - 2} \right\rangle _{\rm{Atom}}$. What we are interested in is the first case $\left[ c'_1  \left| {2G - 2} \right\rangle _{\rm{Atom}}  \left| 1 \right\rangle _{{\rm{AS2}}} \right]$, where the empty spin wave in atoms is expected to enhance Raman scattering. The enhancement effect due to the empty spin wave can be tested by detecting the conditional probability $P_{\rm S3|(S1-AS2)}=N_{\rm S1-AS2-S3}/ N_{\rm S1-AS2}$, where $N_{\rm S1-AS2}$ is coincidence counts of S1 and AS2.  $N_{\rm S1-AS2-S3}$ is the three-fold coincidence counts of S1, AS2 and S3. 

\begin{figure}[th]
\centering 
\includegraphics[width=0.72\textwidth]{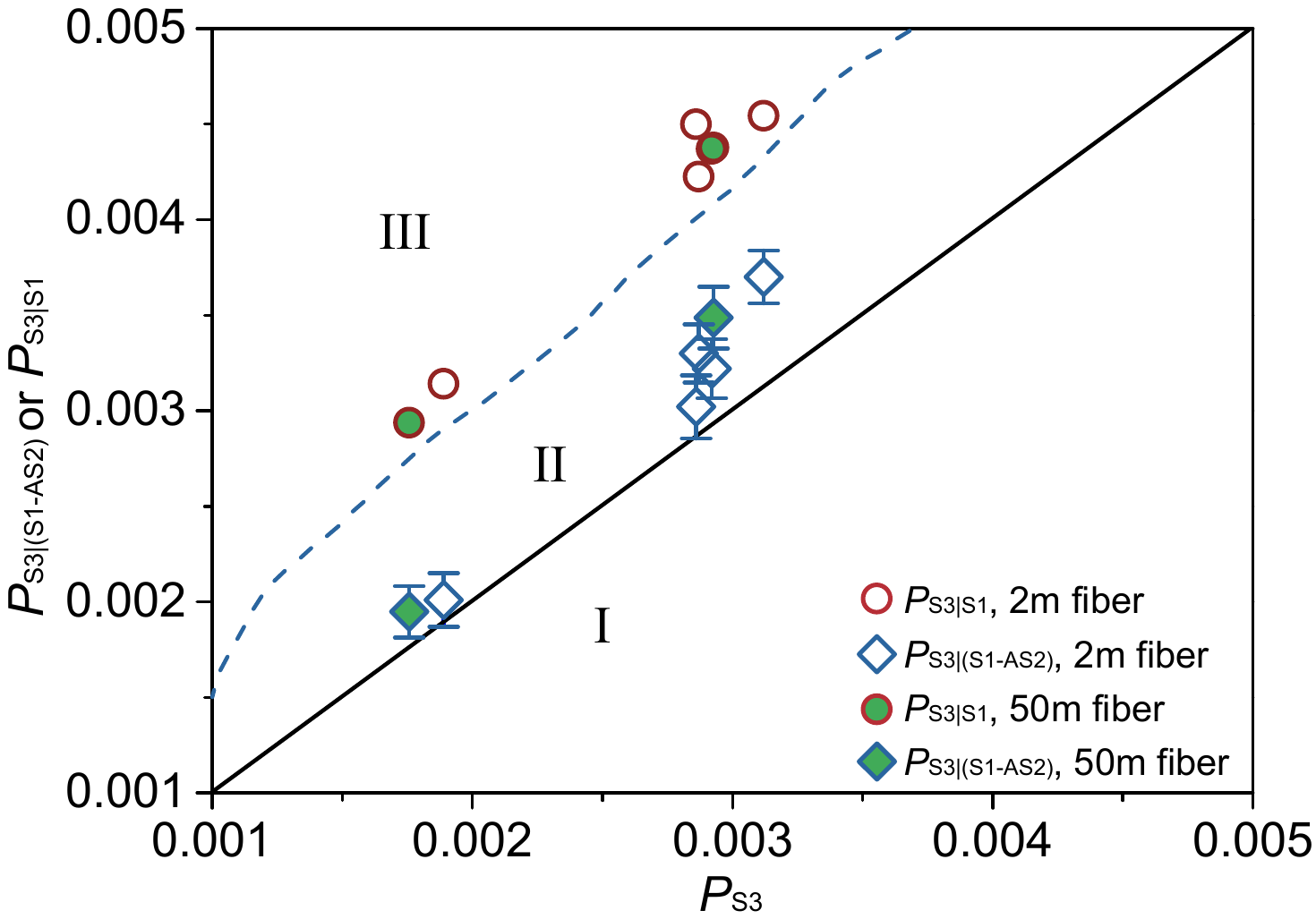}
\caption{\textbf{Distinguishability between a wavefunction $\left| {1} \right\rangle _{\rm{Atom}}$ and a wavefunction $\left| {0} \right\rangle _{\rm{Atom}}$.} $P_{\rm S3}$ is the detected probability of Stokes photons (S3) come from spontaneous Raman scattering due to the probe pulse. $P_{\rm S3|S1}$ is the detected probability of S3 heralded by S1.  $P_{\rm S3|(S1-AS2)}$ is the detected probability of S3 conditioned on the joint detection of S1 and AS2. The filled circles and filled diamonds correspond to the delayed-type experiment where the detection of anti-Stokes photons is delayed by a 50\,m fiber. The blue dashed line is to guide eye and is not a theoretical fitting curve. The solid black line with an angle of 45 degrees depicts the probability values equal to $P_{\rm S3}$. The error bar denotes one standard deviation. The repetition frequency of the experimental trials is 500\,KHz, and the total count time is several hours for each point.}
\label{fig2}
\end{figure}

In this experiment, the retrieval efficiency ranges from 10\% to 20\%, which depends on the energy of read pulse. After the retrieval operation, 80\% to 90\% of the wavefunction is still stored in atoms, no matter whether AS2 is retrieved out or not. And, based on the above assumption that the stored empty spin wave can enhance the Raman scattering as a full wave, the atomic excitations (accompanied by S3) excited by the probe pulse should be independent of the retrieval of AS2. And, the detection probability of S3 should be equal for both cases in (\ref{Bohm02}).  However, the experiment results in Fig.\,\hyperref[fig2]{2} draw a different conclusion. The values of $P_{\rm S3|(S1-AS2)}$ locate in region II are obviously smaller than that of $P_{\rm S3|S1}$ locate in region III. $P_{\rm S3|S1}$ is the sum of the probability of both two cases in (\ref{Bohm02}).

The difference between $P_{\rm S3|(S1-AS2)}$ and $P_{\rm S3|S1}$ lies in whether AS2 is definitively detected. A successful detection of AS2 indicates that the probed spin wave is an empty one. We observe an anti-correlation between AS2 and the enhancement for S3, meaning that a successful detection of AS2 results in the absence of enhanced Raman scattering. Therefore, our results show that the empty spin wave does not provide the same enhancement effect as a full wave. In Fig.\,\hyperref[fig2]{2}, $P_{\rm S3|(S1-AS2)}$ is almost equal to $P_{\rm S3}$ when the probability value $P_{\rm S3}$ is around 0.18\%, which corresponds to an intrinsic excitation probability around 2.5\%. However, $P_{\rm S3|(S1-AS2)}$ is higher than $P_{\rm S3}$ when the excitation probability is high. We attribute this to two factors. The first factor is the indirect enhancement effect following the path $P_{\rm S2|S1}$$ \to $$P_{\rm S3|S2}$. The second factor is the high-order excitation induced by intense control pulses. More details are presented in Appendix D.

The detection of AS2 may lead to collapse of the empty spin wave through some nonlocal effects. For avoiding this nonlocal and instantaneous collapse, we perform a complementary experiment, in which we use a 50\,m fiber instead of a 2\,m fiber to delay the detection of AS2. This long fiber ensures that the the detection of S3 is 55\,ns earlier than the detection of AS2. As is shown in Fig.\,\hyperref[fig2]{2}, the values (filled circles and filled diamonds) of $P_{\rm S3|(S1-AS2)}$ locate in region II are still obviously smaller than that of $P_{\rm S3|S1}$ locate in region III. The delayed-type experiment still shows that the empty spin wave stored in atoms doesn't provide an enhancement effect as a full wave provides.

For a comparison, we further explain above results by the orthodox interpretation of quantum mechanics. The wavefunctions in (\ref{Bohm01}) should be rewritten as a supperposition state
\begin{equation}\label{ampE0}
c_1 \left| {0} \right\rangle _{\rm{Atom}} \left| 1 \right\rangle _{{\rm{AS2}}}  + c_2 \left|1 \right\rangle _{\rm{Atom}} \left| 0 \right\rangle _{{\rm{AS2}}} ,
\end{equation}
rather than two conditional functions. If  there is no 50\,m fiber delay, then AS2 is detected before the probe operation, and the state of (\ref{ampE0}) collapses to $\left| {0} \right\rangle _{\rm{Atom}}$, and no enhancement for S3. On the contrary, in the delayed-type experiment, the detection of S3 is 55\,ns earlier than the detection of AS2, see Fig.\,\hyperref[fig1]{1\,(d)}. Before the detection of AS2, the photon S3 has been translated into a classical information, such as an electric impulse. Detecting a S3 will project the wavefunction, involving atomic excitation and AS2, into the following state
\begin{equation}\label{ampE}
c'_1 \left| {1} \right\rangle _{\rm{Atom}} \left| 1 \right\rangle _{{\rm{AS2}}}  + c'_2 \left| 2 \right\rangle _{\rm{Atom}} \left| 0 \right\rangle _{{\rm{AS2}}} .
\end{equation}
$c'_1$ and $c'_2$ are proportional coefficients dependent of $c_1$, $c_2$ and the enhancement for S3. There is only one atomic excitation in $ \left| {1} \right\rangle _{\rm{Atom}}$, while there are two atomic excitations in $\left| 2 \right\rangle _{\rm{Atom}}$ consisting of a pre-existing atomic excitation heralded by S1 and an atomic excitation heralded by S3.

\begin{figure}[th]
\centering 
\includegraphics[width=0.72\textwidth]{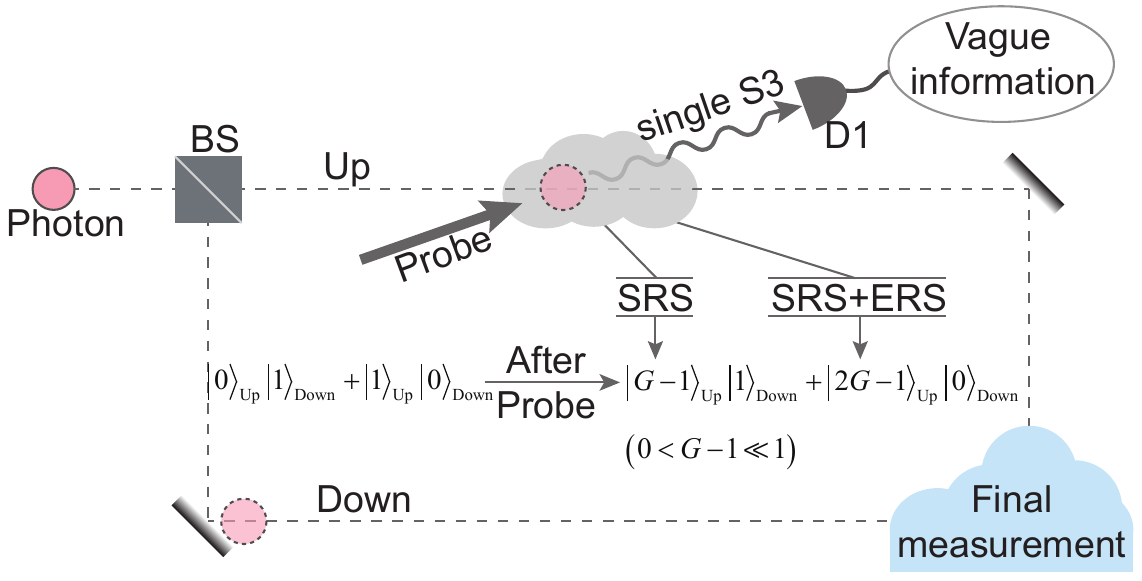}
\caption{\textbf{An analogy for illustrating the probe process.} BS and upper path correspond to the quantum memory, while the lower path corresponds to the propagation path of the flying AS2. The photon depicted by a red circle represents either the atomic excitation (in the upper path) or the flying AS2 (in lower path). BS: beam splitter; SRS: spontaneous Raman scattering; ERS: Raman scattering enhanced by a preexisting atomic excitation; D1: single-photon detector for S3. The generation probability of S3 from Raman scattering is very low ($0<G-1\ll 1$), due to which no which-way information can be obtained by detecting a single photon S3. Here, the average number $G-1$ and $2G-1$ of atomic excitations are used to denote the state of quantum memory after a trial of probing but before the detection of S3, with the  coefficients omitted for simplicity. After a successful detection of S3, the superposition state changes to the state (\ref{ampE}). The wavefunction remains in the form of superposition state until a final measurement forces it into a projection. Note that this is just an analogy, in practice, there is no Raman scattering in an all-optical Mach-Zehnder interferometer.}
\label{fig3}
\end{figure}

As is illustrated in Fig.\,\hyperref[fig3]{3}, the probing of the stored spin wave does not collapse the wavefunction, distinguishing it from typical which-way detectors used in general experiments. This characteristic ensures the survival of the superposition state (\ref{ampE}) after the detection of S3, which is verified by a Mach-Zehnder experiment (see Fig.\,\hyperref[fig7]{7} in Appendix E). An interference visibility of (79$\pm$2)\%, as demonstrated in Fig.\,\hyperref[fig8]{8}, is heralded by the joint detection of S1 and S3. A straightforward explanation for the survival of superposition state is as follows: Each term in the superposition state (\ref{ampE}) corresponds to the generation of a Stokes photon S3 with a low probability. The probe pulse merely provides a small enhancement ($G-1 \ll 1$) to the stored wavefunction. Therefore, `which-way' information cannot be obtained prior to a postselection (i.e. the detection of AS2). It is important to note that if $G-1>1$, the probing of the stored wavefunction would act as a definitive `which-way' detector. This scenario arises because the two terms in (\ref{ampE}) could be distinguished by detecting differences in photon number or intensity of S3, rendering the above analysis invalid. Fortunately, the intrinsical probability of Raman scattering is inherently low in this light-matter interface, making it impossible to distinguish 'which-way' information by detecting a single S3. The two terms in (\ref{ampE}) behaves differently in enhancing the generation of S3, due to which the position information of the atomic excitation hides in the generation probability of S3. However, the hidden position information cannot be revovered prior to postselection on AS2.

In the following, we recount the influences of the probe pulse, as well as the measurement of S3, on the wavefunction of atomic excitation and AS2. After the probe pulse, but before the detection of S3, the states in (\ref{Bohm02}) are rewritten as a superposition state:
\begin{equation}\label{ampE3}
\begin{aligned}
&c_1\left| {G - 1} \right\rangle _{\rm{S3}}\left| {G - 1} \right\rangle _{\rm{Atom}}  \left| 1 \right\rangle _{{\rm{AS2}}}+\\
&c_2 \left| {2G - 2} \right\rangle _{\rm{S3}}\left| 2G-1 \right\rangle _{\rm{Atom}}  \left| 0 \right\rangle _{{\rm{AS2}}} .
\end{aligned}
\end{equation}
S3 is integral to the entire quantum superposition state. Its generation probability, such as average numbers $G-1$ or $2G-2$, correlates to the absence or presence of a preexisting atomic excitation. Detecting a S3 will project the above wavefunction (\ref{ampE3}) onto the state (\ref{ampE}). Due to the enhancement effect from the second term, $|c'_1/c'_2|^2$ is smaller than $|c_1/c_2|^2$. Consequently, the conditional probability of detecting AS2 decreases when S3 is detected. Conversely, if no S3 is detected, the state (\ref{ampE3}) reverts to (\ref{ampE0}) with a high probability close to 100\%, maintaining the probability of generating AS2 as $|c_1|^2$ rather than $|c'_1|^2$. This explains why $P_{\rm S3|(S1-AS2)}$ is smaller than $P_{\rm S3|S1}$ in the delayed-type experiment. It is important to note that a successful detection of S3 converts photon S3 into classical information, such as an electric impulse. Therefore, the observed anti-correlation $P_{\rm S3|(S1-AS2)} < P_{\rm S3|S1}$ suggests that the strong measurement of S3 records vague position information of atomic excitation in a classical manner, thereby nonlocally projecting the quantum wavefunction (\ref{ampE3}) into a new one (\ref{ampE}), rather than the probe pulse directly influencing the atoms. If we were to assume that the probe pulse alone projects the quantum wavefunction (\ref{ampE0}) into (\ref{ampE}), regardless of whether S3 is detected, then the detection probability of AS2 will be independent of S3 detection. However, this assumption contradicts our experimental results. Thus, detecting some energy carried by S3 is crucial for quantum measurement. Of course, the probe pulse itself is not negligible: it generates photon S3 with a low probability, conecting the wavefunction (\ref{ampE0}), involving atomic excitation and AS2, to the photon detector APD1 responsible for the strong measurement of S3.

\section*{3 Discussion}
In summary, our results disprove the notion that an empty spin wave can effectively interfere with other photons in Raman scattering. The empty spin wave appears to be non-interfering when probed in isolation, prompting researchers to seek more plausible interpretations. For example, in 2017, Yakir Aharonov {\it et al.} proposed such an interpretation by virtue of Heisenberg picture \cite{Aharonov2017}: Instead of a quantum wavefunction passing through both slits, the particle goes through only one slit in each trial, but nonlocally interacts with the other slit. A complete description of a quantum system must take into account two boundary conditions: initial state and final state. Another earlier example is from 1967, when L. Mandel {\it et al.} demonstrated an interference between two independent light sources even only one photon is emitted at a time from the two sources \cite{Pfleegor1967, Paul1986}, which Mandel associated with the detection process and the indistinguishability of the paths or sources. In essence, single-photon interference involves two wavefunctions (or states), as well as indistinguishable propagation paths and the measurement device. The hybrid configuration for quantum interference resonates with quantum contextuality \cite{Qu2021, Griffiths2019, Howard2014}, and the complete experimental situation must be considered when describing a quantum interference.

By measuring the coincidence probability of AS2 and S3, we demonstrate a reconciliation of wave-particle duality and quantum nonlocality. In the experiment, the detection of S3 nonlocally projects the superposition state (\ref{ampE0}) onto (\ref{ampE}), preserving the wave nature.  On the other hand, the enhancement for S3 occurs only if the atomic excitation has not been read out (i.e. no AS2 is generated), where the particle nature manifests. Our delayed-type experiment is a new counterpart of delayed-choice experiments \cite{Kim2000, Jacques2007, Peruzzo2012, Kaiser2012, Shadbolt2014, Ma2016, Chaves2018, Wang2019, Yu2019}, suggesting that we seemingly alter an earlier event --- whether the generation of S3 is enhanced or not --- by a later measurement of AS2. In some sense, our delayed-type experiment also serves as a new counterpart to weak measurement experiments, with the key difference being that, in our experiment, the weak value (S3) is detected and recorded before the postselection on AS2. In our hybrid interface, one single quantum is shared by two kinds of objects: caesium atoms and a flying photon. This nonclassical memory-enabled hybrid interface, accompanying with nontrivial probe approaches, may advance fundamental tests of quantum mechanics, such as quantum nonlocality \cite{Dunningham2007, Brunner2014, LiuDN2022, Ringbauer2016, Hao2022, Li2021, Wang2022, Pang2023},  quantum measurement and wave-particle duality separation \cite{Pan2023, Pan2020, Lijiakun2023}.

\medskip
\subsection*{Acknowledgements.}
The authors thank Hui-Jun Li, Yi-Ming Pan, Xiang Wang, Sen Lin and M. Saeed for helpful discussions, and thank Xiao-Ying Shen and Zhe-Yong Zhang for assistance in formatting the figures. This research is supported by the National Key R\&D Program of China (Grants No. 2019YFA0308703, No. 2019YFA0706302, and No. 2017YFA0303700); National Natural Science Foundation of China (NSFC) (Grants No. 62235012, No. 11904299, No. 61734005, No. 11761141014, and No. 11690033, No. 12104299, and No. 12304342);  Innovation Program for Quantum Science and Technology (Grants No. 2021ZD0301500, and No. 2021ZD0300700); Science and Technology Commission of Shanghai Municipality (STCSM) (Grants No. 20JC1416300, No. 2019SHZDZX01, No.21ZR1432800, and No. 22QA1404600); Shanghai Municipal Education Commission (SMEC) (Grants No. 2017-01-07-00-02-E00049); Startup Fund for Young Faculty at SJTU (SFYF at SJTU) (Grants No. 24X010500170); China Postdoctoral Science Foundation (Grants No. 2020M671091, No. 2021M692094, No. 2022T150415). X.-M.J. acknowledges additional support from a Shanghai talent program and support from Zhiyuan Innovative Research Center of Shanghai Jiao Tong University.
 
\subsection*{Competing Interests}	
The authors declare that they have no competing financial or non-financial interests.

\subsection*{\bf Data availability.}
The data that support the findings of this study are available from the corresponding authors on reasonable request.

\medskip
\section*{A. Details of the experimental setup.}
The $^{133}$Cs atoms are sealed in a glass cell with a length of 75\,mm. The glass cell is packed in a three-layer magnetic shielding and is heated up to 61\,$^{\circ}$C. A three-level $\Lambda$-type configuration is adopted. The lower two energy states are $\left | 1 \right \rangle \left( 6S_{1/2}, F=3\right)$ and $\left | 2 \right \rangle\left( 6S_{1/2}, F=4\right)$, {\it i.e.} the hyperfine ground states of $^{133}$Cs, and the excited state $\left | 3 \right \rangle$ is the $ 6P_{3/2}$ manifold. The initial state refers to the state that almost all atoms populate on the energy level $\left | 1 \right \rangle$. The pump light for preparing atomic initial state comes from an external cavity diode laser, and is resonant with the transition $\left( 6S_{1/2}, F=4 \to  6P_{3/2} \right)$. The pump light is turned on at least 1\,$\mu{\rm s}$ before each attempt and is turned off about 100\,ns before the arriving of the first control pulse. 

We utilize a set of control pulses that come from the same one distributed Bragg reflector (DBR) laser, which ensures a stable phase relation between the Stokes photons, the atomic excitation and the retrieved anti-Stokes photons in the test duration about 400\,ns. A fast electro optic modulator (EOM) triggered by electronic pulses from an arbitrary waveform generator is used to chop the DBR laser to short pulses with tunable amplitudes. The generated pulses are fed into a tapered amplifier (TA) to boost their power. In order to eliminate the spontaneous emission from the TA, we employ a ruled diffraction grating to spread beam out and spatially pick the stimulated radiation with irises. The control pulses are horizontally polarized by a Glan-Taylor polarizer. In this letter, the control pulse width is 4\,ns, and the beam waist in the glass cell is about 350\,$\mu$m. For measuring the data in Fig.\,2 in the main text, the pulse energy of the write pulse and probe pulse is set to 220\,pJ (400\,pJ) for obtaining lower (higher) excitation probability. For the data around $P_{\rm S3} = 0.003$, several energy values of the read pulse including 220\,pJ, 400\,pJ and 500\,pJ are applied in order to verify that applying a different read pulse energy doesn't affect the conclusion that $P_{\rm S3|(S1-AS2)}$ is smaller than $P_{\rm S3|S1}$. In our experiment, the retrieval efficiency ranges from 10\% to 20\%. For avoiding the resonant fluorescence noise, the control pulses are red detuned ${3.87\,{\rm GHz}}$ from the transition $\left( 6S_{1/2}, F=4 \to 6P_{3/2}, F=4 \right) $.

The Stokes photons and retrieved anti-Stokes photons possess a same polarization vertical to that of control pulses \cite{Nunn2008, Dou2018}. Here, we use a Wollaston prism (not shown in Fig.\,1 in the main text) to basically filter out them from the control pulses. There are still too much noise just after the Wollaston prism due to the leakage or scattering of control pulses. In our experiment, we use a home-made frequency filter (cascaded cavities) to filter out noise photons and separate the Stokes photons and anti-Stokes photons. The peak transmission of each cavity is higher than 90\%. The peak transmission of the whole filter is about 70\%, and the extinction ratio is about $10^{7}$.  The full width at half maximum of the total transmission window is 380\,MHz. 

After passing through the filter, the Stokes (anti-Stokes) photons are collected by a collimator and a single-mode fiber with a length of 2\,m (or 50\,m). Then, the photons are transmitted to single-photon detectors, i.e. avalanche photo-diodes. The time interval between the control pulses is about 150\,ns, which ensures that the after pulses from the single-photon detector \cite{Liu2022, Hu2022} almost disappear before the arrival of subsequent photons. The electric signals output from the detectors are sent to a multi-channel counting system where the arriving time of each signal is recorded.

\begin{figure}[th!]
\centering
\includegraphics[width=0.66\textwidth]{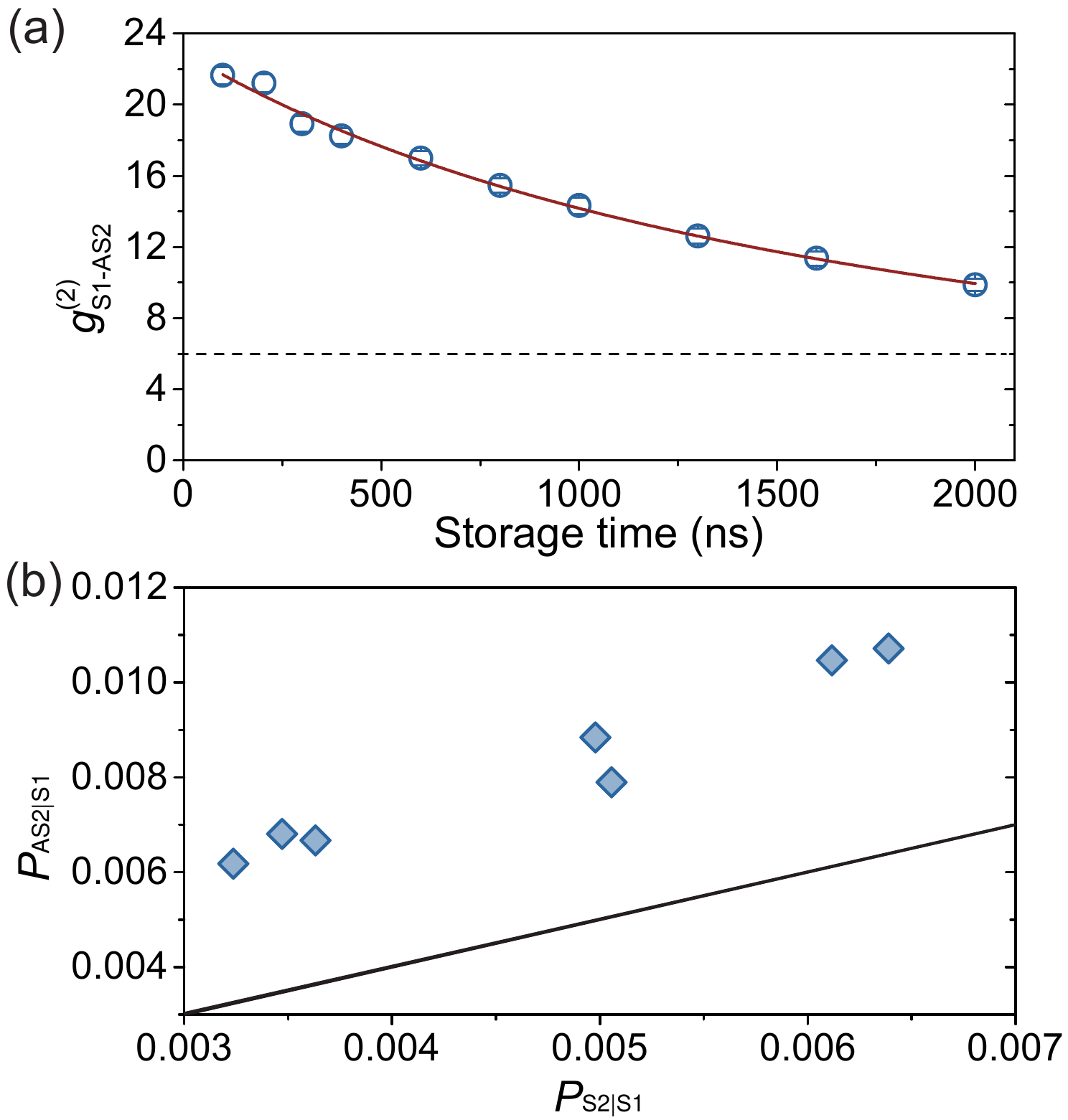}
\caption{\textbf{Cross correlation and heralding efficiency of the light-matter interface.} ({\bf a}) The cross correlation of  S1 and AS2 as a function of storage time. The red curve is a theoretical fit of the form $g_{\rm S-AS}^{(2)}=1+C/(1+At^2+Bt)$.  ({\bf b}) The detected retrieval probability $P_{\rm AS2|S1}$ and the detected excitation probability $P_{\rm S2|S1}$ of S2 heralded by S1, where S2 denotes Stokes photons generated by the second control pulse, i.e. the read pulse. $P_{\rm AS2|S1}$ is higher than $P_{\rm S2|S1}$, especially in consideration of the total detection efficiency of Stokes (anti-Stokes) photons about 7.1\% (5.4\%). The solid balck line depicts the probability values equal to $P_{\rm S2|S1}$. In order to present the raw probability values, each probability has not been divided by its corresponding total detection efficiency. } 
\label{fig4}
\end{figure}

\section*{B. Cross correlation and efficiency of the quantum memory enabled light-matter hybrid interface.}
Before testing the enhanced Raman scattering, we firstly characterize the performance of the hybrid interface, and the results are shown in Fig.\,\hyperref[fig4]{4}. Our interface works in quantum regime (cross correlation higher than 2), see Fig.\,\hyperref[fig4]{4a}. The probing of stored spin wave is performed within a storage time interval (0, 420\,ns). In this time interval, the cross correlation is around 20 higher than the boundary value 6 for violating Bell's inequality. This verifies the quantum nature of our experiment, since a cross correlation higher than 2 means a sub-Poissonian distribution of the heralded AS2 as well as the heralded atomic excitation. For a classical light, such as a coherent light or a thermal light, the cross correlation is not higher than 2. In Fig.\,\hyperref[fig4]{4b}, the retrieval probability $P_{\rm AS2|S1}$ is higher than the excitation probability $P_{\rm S2|S1}$ of S2 heralded by S1, especially in consideration of the total detection efficiency of Stokes (anti-Stokes) photons about 7.1\% (5.4\%). In order to present probability values really detected, the probability values in this letter have not been divided by the corresponding total detection efficiency, unless otherwise specified. As $P_{\rm AS2|S1}$ is higher than $P_{\rm S2|S1}$, after the successful retrieval of AS2, the probability of atomic excitation in atoms is much decreased even though $P_{\rm S2|S1}$ may provide some gain (smaller than 0.1) to the atomic excitation correlated with S1. Therefore, we are sure that the probability of a full wave remained in atoms is much decreased after the successful retrieval of AS2.

\section*{C. Enhanced Raman scattering due to a preexisting atomic excitation.} 

\begin{figure}[ht]
\centering 
\includegraphics[width=0.7\textwidth]{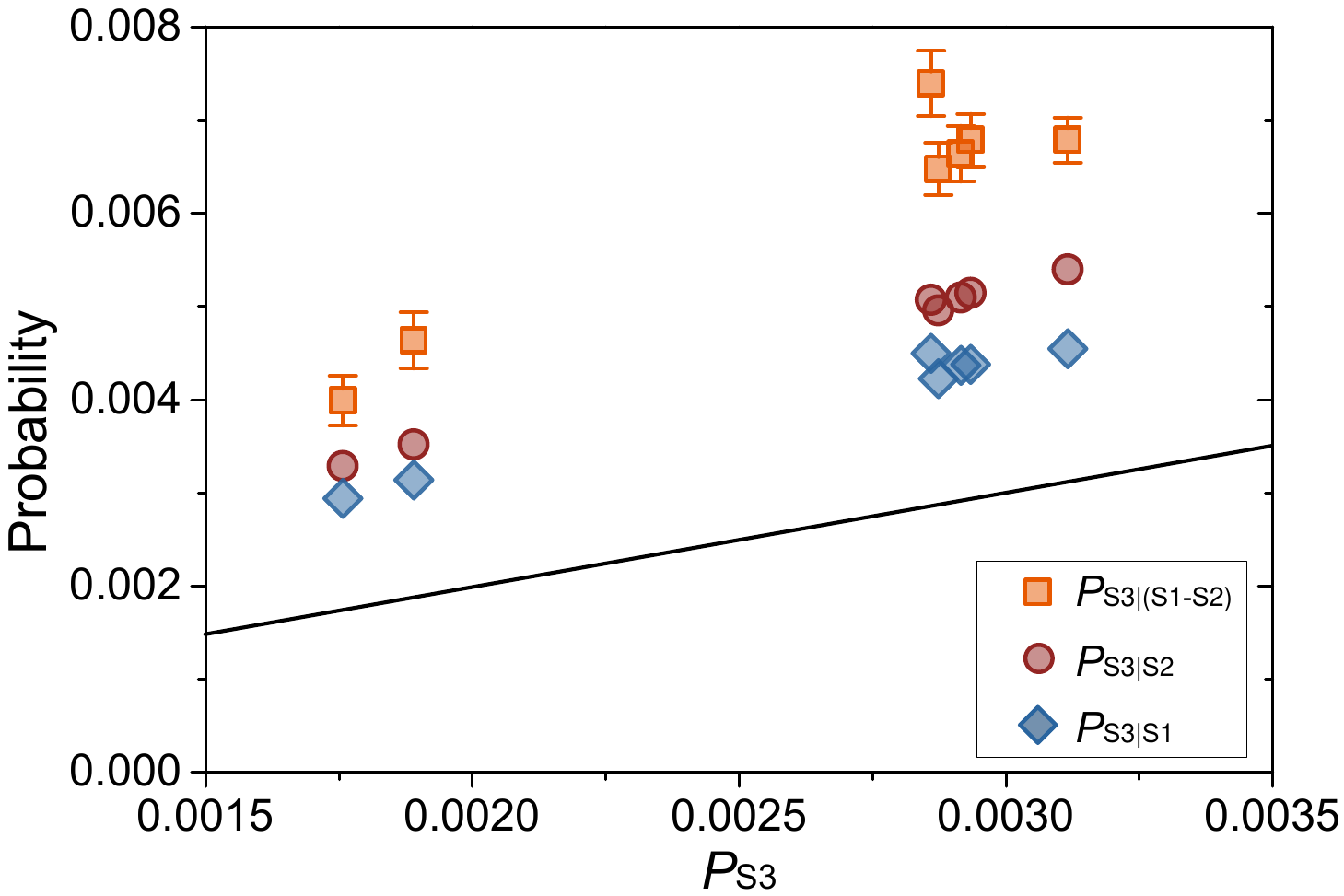}
\caption{\textbf{Enhancement effect due to preexisting atomic excitations. } $P_{\rm S3|(S1-S2)}$ denotes the detected probability of S3 heralded by the joint detection of S1 and S2. Three kinds of heralded probability are used to show that more preexisting atomic excitations provides higher enhancement effect. The solid black line depicts the probability values equal to $P_{\rm S3}$.  }
\label{fig5}
\end{figure}
Assume that a probe pulse with a square waveform and a width $\delta t$ enters the atomic ensemble at time $t=0$, meanwhile there are $n_{{\rm{S}}}(t=0)$ Stokes photons and $n_{{\rm{E}}}(t=0)$ atomic excitations in the atomic ensemble. Then an amplification process for $n_{{\rm{S}}}$ and $n_{{\rm{E}}}$ will happen, which is induced by the probe pulse, and is enhanced by the preexisting seminal Stokes photons and atomic excitations \cite{Louisell1961,Chen2009}. At a time $t$  ($0 \leq t \leq \delta t$), the average number of Stokes photons $ \bar n_{\rm{S}} (t)$ and atomic excitations $\bar n_{\rm{E}} (t)$ are given by
\begin{equation}\label{eqRaman}
\begin{array}{l}
 \bar n_{\rm{S}} (t) =  n_{{\rm{S}}}(0) \cosh ^2 (\kappa t) + \left( {1 +  n_{{\rm{E}}}(0) } \right)\sinh ^2 (\kappa t) , \\ 
 \bar n_{\rm{E}} (t) =  n_{{\rm{E}}}(0) \cosh ^2 (\kappa t) + \left( {1 +  n_{{\rm{S}}}(0) } \right)\sinh ^2 (\kappa t) , \\ 
 \end{array}
\end{equation}
where $\kappa$ is a coupling coefficient. In this letter, there is no Stokes photon in the atomic ensemble at $t=0$, i.e. $ n_{{\rm{S}}}(0) =0 $. By substituting $\left [ \sinh ^2 (\kappa t) = \cosh ^2 (\kappa t) - 1\right]$ into Eq.(\ref{eqRaman}), we can obtain the average number of Stokes photons at time $t$,
\begin{equation}
\bar n_{\rm{S}} (t) =  {n_{{\rm{E}}}(0)  } \left[ {\cosh ^2 (\kappa t) - 1} \right] + \left[ {\cosh ^2 (\kappa t) - 1} \right] .
\end{equation}
Note that the last term $\left[ {\cosh ^2 (\kappa t) - 1} \right]$ corresponds to the photon number due to spontaneous Raman scattering. In this letter, there is one atomic excitation stored in the atomic ensemble, i.e. ${n_{{\rm{E}}}(0) }=1 $, thus $\bar n_{\rm{S}} (t)$ will be $2\left[ {\cosh ^2 (\kappa t) - 1} \right]$ which is twice of the spontaneous Raman scattering. In other words, the Raman scattering is enhanced by the preexisting atomic excitation. It is this enhancement effect that we will use to probe the effect of empty spin wave and full spin wave in this letter. It is worth mentioning that, in practice, $P_{\rm S3|S1}$ is smaller than the twice of $P_{\rm S3|(S1-AS2)}$ as is shown in Fig.\,2 in the main text, since there is a non-negligible probability that the atomic excitation heralded by S1 decays before the arrival of probe pulse or is retrieved out by the read pulse.

In Fig.\,\hyperref[fig5]{5}, we present three kinds of heralded probability. We can see that the preexisting atomic excitation can certainly enhance the generation of S3. More preexisting atomic excitations provides higher enhancement effect. $P_{\rm S3|S1}$ is slightly smaller than $P_{\rm S3|S2}$, which we attribute to the slow decay of the atomic excitation in atoms and the probabilistic retrieval of AS2 by the read pulse.

\begin{figure}[th]
\centering 
\includegraphics[width=0.7\textwidth]{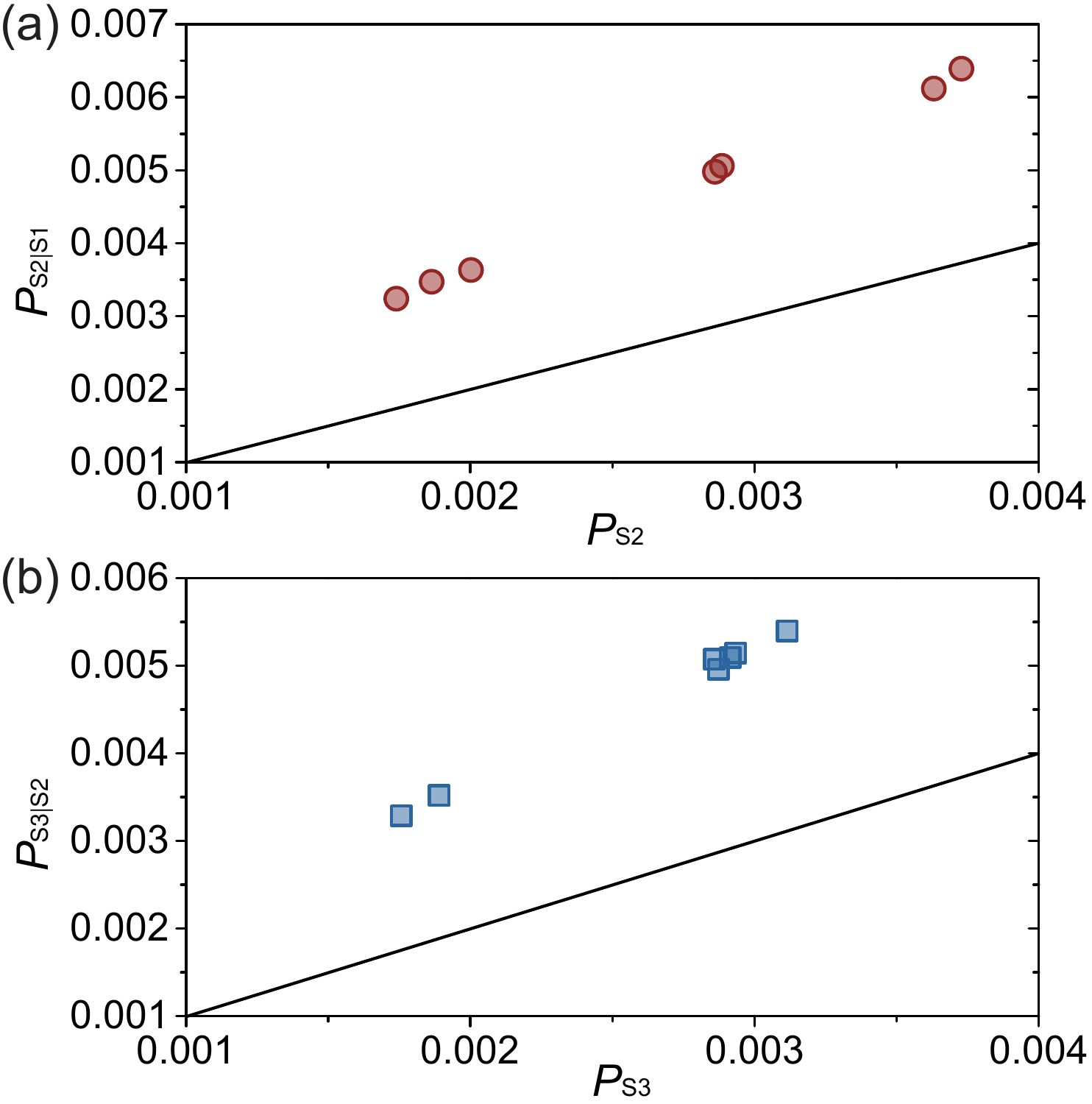}
\caption{\textbf{Excitation probability enhanced by a preexisting atomic excitation. } (a) S2 is enhanced by the atomic excitation that corresponds to S1. (b) S3 is enhanced by the atomic excitation that corresponds to S2. Following the path $P_{\rm S2|S1}$$ \to $$P_{\rm S3|S2}$, S3 experiences an indirect enhancement that originates from the atomic excitation correlated with S1. The error bars are too small.}
\label{fig6}
\end{figure}
\section*{D. Indirect enhancement effect and high-order excitations.} 
In Fig.\,\hyperref[fig6]{6a}, S2 is enhanced by the preexisting atomic excitation that corresponds to the heralding Stokes photon S1. In Fig.\,\hyperref[fig6]{6b}, S3 is enhanced by the atomic excitation that corresponds to S2 which is enhanced by the preexisting atomic excitation corresponds to S1. Thus, S3 experiences an indirect enhancement by the atomic excitation corresponds to S1.

When intense control pulses are used, the probability of high-order excitations can not be ignored. For example, in Fig.\,\hyperref[fig6]{6a}, the maximum probability of $P_{\rm S2|S1}$ is 0.64\%. Taking into account the total detection efficiency 7.1\% of Stokes photons, the intrinsic heralded probability of the atomic excitation corresponds to $P_{\rm S2|S1}$ is about 9\%. Thus, the average number of the atomic excitations due to high-order excitation is about $2\times 0.09^2=0.016$, which is 18\% of 0.09. The high-order excitations due to intense control pulses will stay in the atoms with a high probability even though one AS2 is retrieved out, and then enhance the scattering of S3.

\section*{E. Interference between two read-out modes of anti-Stokes photons.} 
\begin{figure*}
\centering 
\includegraphics[width=0.96\textwidth]{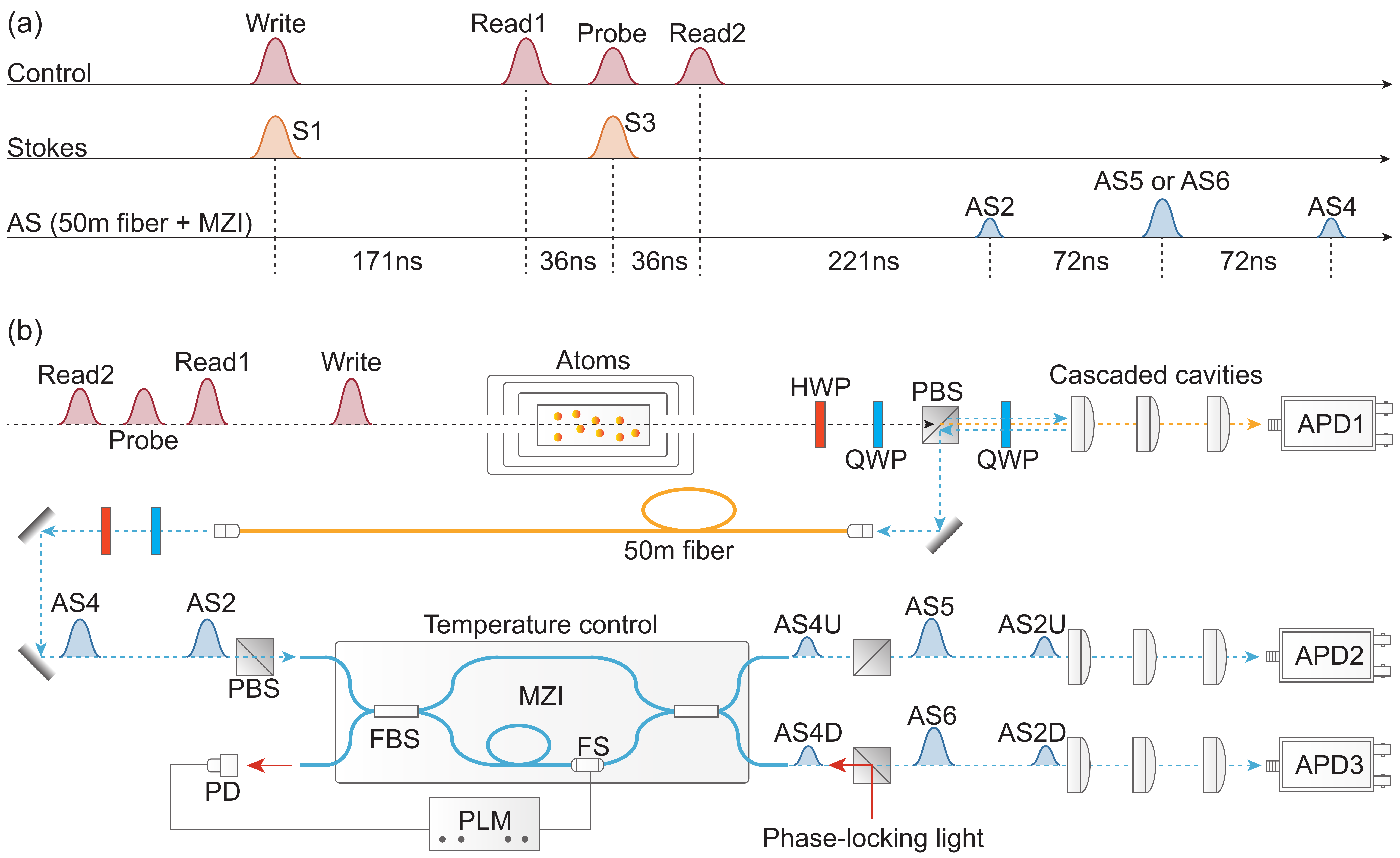}
\caption{\textbf{Experimental demonstration of the survival of the superposition state after the detection of S3. } (a) Time sequence. AS2 is read out by the first read pulse (second control pulse), and AS4 is read out by the second read pulse. AS5 and AS6 consist of AS2 passed through the long arm of Mach-Zehnder interferometer (MZI) and AS4 passed through the short arm of MZI. The detection of anti-Stokes photons is delayed by a 50\,m fiber and the MZI. (b) Experimental setup. FBS: fiber beam splitter. PLM: phase locking module. PD: photo-diode. FS: fiber stretcher. The temperature fluctuation of the MZI is locked within $\pm$1\,mK. The temperature fluctuation of the cascaded cavities is locked within $\pm$2\,mK. The phase-locking light and the control light come from a same one distributed Bragg reflector (DBR) laser. The power of phase-locking light is 540\,nW. The phase difference $\beta$ between $\psi(\vec r, t)_{\rm AS2}$ passed through the long arm and $\psi(\vec r, t)_{\rm AS4}$ passed through the short arm can be tunned by slightly changing the temperature of the MZI, when the phase-locking light is all the time locked at its peak or trough of transmission detected by the PD. Note that there are about 14\,m-length difference between the two arms of the MZI, and the changes of phase difference $\beta$ for horizontally polarized anti-Stokes photons and vertically polarized phase-locking light are different as the temperature changes due to the large birefringence of the polarization-maintaining fiber. Stokes photons are detected by APD1. Anti-Stokes photons are detected by APD2 or APD3. The letter `U' in AS2U denotes that the anti-Stokes photons are detected by APD2, and `D' for those detected by APD3.}
\label{fig7}
\end{figure*}
In the following, we start from an analysis based on the joint probability distribution of photons S3 and AS. After the probe process, if the residual atomic spin wave is retrieved as a wave of anti-Stokes photon by another read pulse (the 4th control pulse) just after the probe pulse as in Fig.\,\hyperref[fig7]{7}, then the quantum state involving S3 and anti-Stokes photon (AS) can be written as
\begin{equation}\label{retrieveE}
\left| {\Psi} \right\rangle \propto c_3\left| {\psi} _{\rm{G-1}} \right\rangle \left| \phi_{\rm AS2} \right\rangle   + c_4 \left| \psi_{\rm{2G-2}} \right\rangle  \left| \phi_{\rm AS4} \right\rangle ,
\end{equation}
where $\left| {\psi} _{\rm{G-1}}\right\rangle$ ($\left| {\psi} _{\rm{2G-2}}\right\rangle$) denotes the wavefunction of Stokes photon S3 with an average number of $G-1$ ($2G-2$) in each trial, and $\left| \phi_{\rm AS2}\right\rangle$ ($\left| \phi_{\rm AS4}\right\rangle$) denotes the wave of anti-Stokes photon AS retrieved by the 2nd (4th) control pulse. $c_4$ depends on the retrieval efficiency and the parameter $c'_2$ in state  (\ref{ampE}). The joint probability distribution of photons S3 and AS at position $\vec r$ and time $t$ can be written as the square modulus of ${\Psi (\vec r, t)}$:
\begin{equation}\label{squareW}
\begin{small}
\begin{aligned}
\left| {\Psi (\vec r, t)} \right|^{\rm 2} & \propto  \left| c_3{\psi(\vec r, t)} _{\rm{G-1}}  \phi(\vec r, t)_{\rm AS2} \right|^{\rm 2}   \\
&+ \left|  c_4 {\psi(\vec r, t)}_{\rm{2G-2}} \phi(\vec r, t)_{\rm AS4} \right|^{\rm 2}\\
&+c_3^*c_4{\psi(\vec r, t)}^* _{\rm{G-1}}{\psi(\vec r, t)}_{\rm{2G-2}}\phi(\vec r, t)^*_{\rm AS2} \phi(\vec r, t)_{\rm AS4}\\
&+c_3 c_4^*{\psi(\vec r, t)} _{\rm{G-1}}{\psi(\vec r, t)}^*_{\rm{2G-2}}\phi(\vec r, t)_{\rm AS2} \phi(\vec r, t)^*_{\rm AS4} \,\,.\\
\end{aligned}
\end{small}
\end{equation}

{\it Case1}: Assume that $\phi(\vec r, t)_{\rm AS2}$ and $\phi(\vec r, t)_{\rm AS4}$ are not overlapped in space and time, and AS2 is detected before the detection of S3, and the superposition state collapses due to the detection of AS2, then the measurement result of none enhanced Raman scattering is exactly predicted by the first term in (\ref{squareW}).

{\it Case2}: On the contrary, in the delayed-type experiment, the detection of S3 is 55\,ns earlier than the detection of AS2, see Fig.2 in the main text. Assume that $\phi(\vec r, t)_{\rm AS2}$ and $\phi(\vec r, t)_{\rm AS4}$ are not overlapped in space and time, then the last two terms are zeros. And, we only need to consider the first two terms. By detecting a singe S3, one cannot distinguish the first two terms in (\ref{squareW}). That is, the detection of a single photon S3 doesn't predict and determine which one of AS2 and AS4 will be detected in a final measurement. Then, how does the anti-correlation ($P_{\rm S3|(S1-AS2)}< P_{\rm S3|S1}$) between AS2 and enhanced S3 occur? Note that  $P_{\rm S3|(S1-AS2)}=N_{\rm S1-AS2-S3}/ N_{\rm S1-AS2}$, which is not $N_{\rm S1-AS2-S3}/ N_{\rm S1}$ or $N_{\rm S1-AS2-S3}/ N_{\rm trial}$, where $N_{\rm trial}$ is the total number of trials. 

{\it Case3}: If $\phi(\vec r, t)_{\rm AS2}$ and $\phi(\vec r, t)_{\rm AS4}$ are overlapped in space and time, then there will be interference effects due to the last two terms in (\ref{squareW}). The interference pattern depends on the phase relations and probability amplitudes of photons S3, AS2 and AS4. It is challenging to observe this interference pattern until the phase relation between these photons is well locked and stabilized. Another intractable point is that AS4 heralded by the joint detection of S1 and S3 is composed of two parts: Firstly, the read out of the atomic excitation generated by the write pulse and heralded by S1; Secondly, the read out of the atomic excitation generated by the probe pulse and heralded by S3. The second part involved in AS4 makes the interference pattern complex. In a word, it is not easy to observe a high-visibility interference between AS2 and AS4, heralded by the joint detection of S1 and S3. 
\begin{figure}[th!]
\centering 
\includegraphics[width=0.7\textwidth]{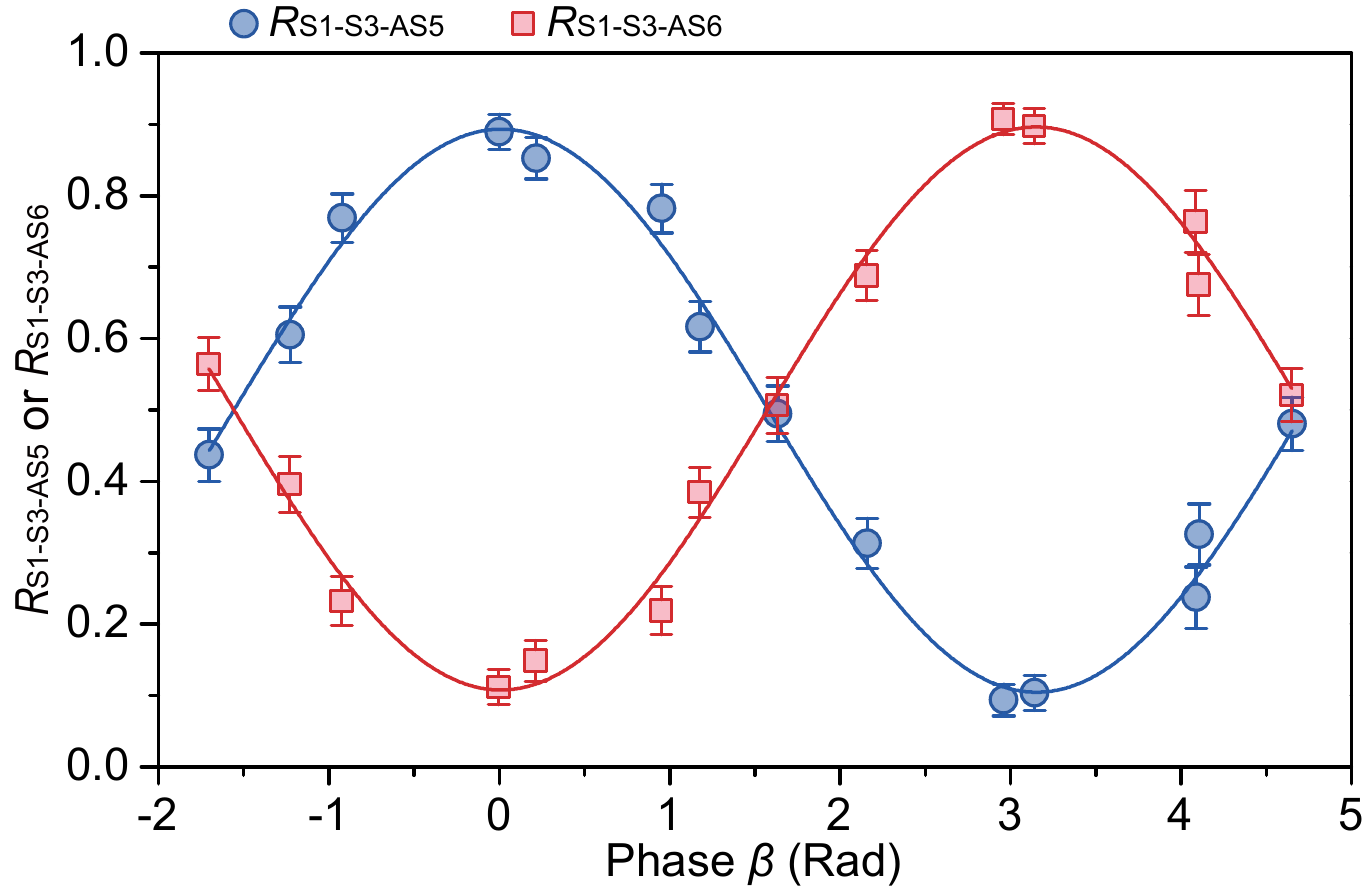}
\caption{\textbf{Interference between two read-out modes of anti-Stokes photons AS2 and AS4, conditioned on the joint detection of S1 and S3.} Blue circles correspond to $R_{\rm S1-S3-AS5}=N_{\rm S1-S3-AS5}/(N_{\rm S1-S3-AS5}+\eta_{2\rm UD} N_{\rm S1-S3-AS6})$. $\eta_{2\rm UD}=N_{\rm S1-AS2\rm U}/N_{\rm S1-AS2\rm D}$ is the ratio of the total detection efficiencies of the two channels of anti-Stokes photons. $N_{\rm S1-S3-AS5}$ denotes the three-fold coincidence counts involving S1, S3 and AS5. Red squares correspond to $R_{\rm S1-S3-AS6}=1-R_{\rm S1-S3-AS5}$. The error bar denotes one standard deviation. The blue curve and red curve are theoretical fits of the form $R=a[0.5+0.5V{\rm cos}(\beta+\delta)]$. The visibility is $V=(79\pm2)\%$. The repetition frequency of the experimental trials is 1\,MHz, and the total count time is 5 hours for each point.}
\label{fig8}
\end{figure}

In order to maintain a continuously stable phase relation of S3, AS2 and AS4, the four control pulses come from the same one distributed Bragg reflector (DBR) laser. The phase difference $\beta$ between AS2 and AS4 in the Mach-Zehnder interferometer (MZI) is locked by a phase-locking module, and the temperature of the MZI is stabilized with a precision of 2\,mK. In addition, for obtaining an interference pattern with a high visibility, we try to balance the amplitudes of heralded AS2 and heralded AS4 by using a set of control pulses with different pulse energies. The pulse energy of write pulse (probe pulse) is 380\,pJ (270\,pJ). The pulse energy of first (second) read pulse is the same to that of the write pulse (probe pulse). The measured visibility $V$ of single-photon interference is up to ($79\pm 2)\%$, as is shown in Fig.\,\hyperref[fig8]{8}.

\section*{F. Estimated visibility if the detection of S3 collapses the superposition state.} 
If our probe method can distinguish the two terms of the superposition state 
\begin{equation}\label{APampE}
c'_1 \left| {1} \right\rangle _{\rm{Atom}} \left| 1 \right\rangle _{{\rm{AS2}}}  + c'_2 \left| 2 \right\rangle _{\rm{Atom}} \left| 0 \right\rangle _{{\rm{AS2}}} .
\end{equation}
without notifying any observer, and the wavefunction of superposition state collapses in the probe process, then there are two possible cases: 

\noindent {\it Case A},  the superposition state collapses to $\phi(\vec r, t)_{\rm AS4}$ corresponding to the second term in the state (\ref{APampE}) , and nothing is left in $\phi(\vec r, t)_{\rm AS2}$ for the survival part (AS4) to interfere with, and the interference pattern with a visibility up to ($79\pm 2)\%$ should not be observed. 

\noindent {\it Case B}, the superposition state collapses to $\phi(\vec r, t)_{\rm AS2}$ corresponding to the first term in the state (\ref{APampE}). There may be an interference pattern due to that the wave of $\phi(\vec r, t)_{\rm AS4}$  (comes from the read out of the atomic excitation generated by the probe pulse and heralded by S3) may interfere with the wave of $\phi(\vec r, t)_{\rm AS2}$ (comes from the read out of the atomic excitation generated by the write pulse and  heralded by S1). That is, we assume that there is an interference between waves come from different atomic excitations. Note that, the coincidence $N_{\rm S1-S3-AS2-AS4}$ of the full wave of AS2 and the full wave of AS4 heralded by the joint detection of S1 and S3 corresponds to a four-fold coincidence, and can be ignored in comparison with three-fold coincidence $N_{\rm S1-S3-AS2}+N_{\rm S1-S3-AS4}$, therefore we only need to consider the interference between the full wave of AS2 heralded by S1 and the wave (including full spin wave and empty spin wave) of AS4 heralded by S3.

Here, we make an estimation of the possible visibility of the overall interference pattern of the above two cases based on the following measured results:
\begin{equation}\label{Vdata}
\begin{aligned}
&P_{\rm S1}=N_{\rm S1}/N_{\rm trial}=0.147\%, \\
&P_{\rm S3}=N_{\rm S3}/N_{\rm trial}=0.126\%, \\
&P_{\rm S3|S1}=N_{\rm S1-S3}/N_{\rm S1}=0.219\%,\\
&P_{\rm AS2|S1}=P_{\rm S1-AS2U}+P_{\rm S1-AS2D}=0.128\%, \\
&P_{\rm AS4|S1}=P_{\rm S1-AS4U}+P_{\rm S1-AS4D}=0.076\%,\\
&P_{\rm AS4|S3}=P_{\rm S3-AS4U}+P_{\rm S3-AS4D}=0.118\%,
\end{aligned}
\end{equation}
where the error bars are too small. Refering to the general definition of visibility
\begin{equation}
V=\frac{I_{\rm max}-I_{\rm min}}{I_{\rm max}+I_{\rm min}}=\frac{2\sqrt{I_{\rm 1}I_{\rm 2}}}{I_{\rm 1}+I_{\rm 2}},
\end{equation}
where $I_{\rm 1}$ and $I_{\rm 2}$ denote the intensities of two beams that interfere with each other, we can estimate the possible visibility $V_{\rm E}$ also heralded by the joint detection of S1 and S3.
\begin{equation}\label{possibleV}
\begin{aligned}
V_{\rm E}&=\frac{2\sqrt{P_{\rm AS2|S1}P_{\rm AS4|S3}}}{|c_{5}|^{2}(P_{\rm AS4|S1}+P_{\rm AS4|S3})+(P_{\rm AS2|S1}+P_{\rm AS4|S3})}\\
&=55\%,
\end{aligned}
\end{equation}
The sum $(P_{\rm AS4|S1}+P_{\rm AS4|S3})$ in the denominator corresponds to the case A in which the superposition state collapses to $\phi(\vec r, t)_{\rm AS4}$ consists of the waves of AS4 heralded by S1 and S3, and nothing is left in $\phi(\vec r, t)_{\rm AS2}$ for the survival part (AS4) to interfere with. The sum $(P_{\rm AS2|S1}+P_{\rm AS4|S3})$ corresponds to the case B in which the superposition state collapses to $\phi(\vec r, t)_{\rm AS2}$, and an interference between a full wave $\phi(\vec r, t)_{\rm AS2}$ heralded by S1 and a wave (including full spin wave and empty spin wave) of $\phi(\vec r, t)_{\rm AS4}$ heralded by S3 may exist. The proportionality coefficient $|c_{5}|^{2}=(P_{\rm AS4|S1}/P_{\rm AS2|S1})(P_{\rm S3|S1}/ P_{\rm S3})$. $(P_{\rm AS4|S1}/P_{\rm AS2|S1})$ denotes the ratio of the retrieval efficiency of the second read pulse and the retrieval efficiency of the first read pulse, which is determined by the configuration (the energy settings) of the two read pulses. $(P_{\rm S3|S1}/ P_{\rm S3})$ denotes the enhanced effect due to the preexisting atomic excitation heralded by S1, note that a successful detection of S3 corresponds to a higher probability for $\left| 2 \right\rangle _{\rm{Atom}} \left| 0 \right\rangle _{\rm{AS2}}$ than that for $\left| 1 \right\rangle _{\rm{Atom}} \left| 1 \right\rangle _{\rm{AS2}}$ in the superposition state (\ref{APampE}). For case A, the atomic excitation survives after the first read pulse, and thus enhances the  generation of S3 by a factor of $(P_{\rm S3|S1}/ P_{\rm S3})$. Conversely, a successful detection of S3 means a gain factor $(P_{\rm S3|S1}/ P_{\rm S3})$ for the probability of {\it case A}.

The estimated visibility $V_{\rm E}=55\%$ is obviously smaller than the observed $79\%$. In practice, the visibility $V_{\rm E}$ will be reduced by the the temperature fluctuation and the unbalanced splitting ratio of the MZI, and a real value of $V_{\rm E}$ should be smaller than $55\%$. Therefore, the explanation that wavefunction collapses after a successful detection of S3 is inadequate of interpreting the observed visibility up to ($79\pm 2)\%$, even though the case in which an identified full wave $\phi(\vec r, t)_{\rm AS2}$ and an identified wave $\phi(\vec r, t)_{\rm AS4}$ may interfere with each other is taken into account. At this point, the survival of the superposition state (\ref{APampE}) after the detection of S3 is verified by the Mach-Zehnder experiment.

\section*{G. Approaches to improve the visibility.} 
\begin{figure}[th]
\centering 
\includegraphics[width=0.7\textwidth]{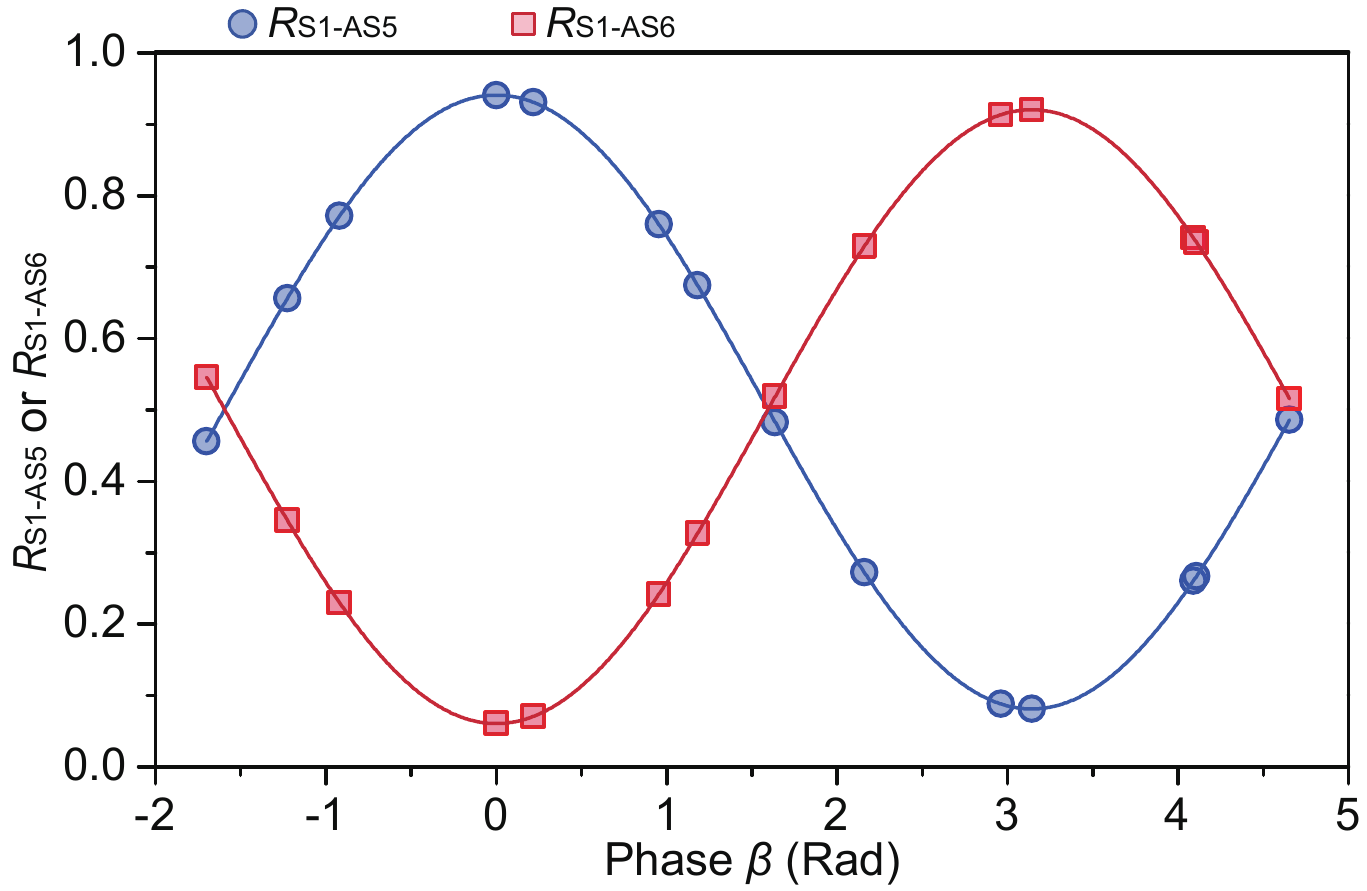}
\caption{\textbf{Interference between two read-out modes of anti-Stokes photons AS2 and AS4, conditioned on the detection of S1.} Blue circles correspond to $R_{\rm S1-AS5}=N_{\rm S1-AS5}/(N_{\rm S1-AS5}+\eta_{2\rm UD} N_{\rm S1-AS6})$. $\eta_{2\rm UD}=N_{\rm S1-AS2\rm U}/N_{\rm S1-AS2\rm D}$ is the ratio of the total detection efficiencies of the two channels of anti-Stokes photons. $N_{\rm S1-AS5}$ denotes the coincidence counts involving S1 and AS5. Red squares correspond to $R_{\rm S1-AS6}=1-R_{\rm S1-AS5}$. The error bars are too small to be visible due to the huge coincidence counts. The blue curve and red curve are theoretical fits of the form $R=a[0.5+0.5V_2{\rm cos}(\beta+\delta)]$. The visibility is $V_2=86\%$, and the standard deviation can be ignored. The repetition frequency of the experimental trials is 1\,MHz, and the total count time is 5 hours for each point.}
\label{fig9}
\end{figure}
The possible approaches to improve the visibility includes optimizing the balance between AS2 and AS4, and suppressing the temperature fluctuation of the MZI. During this experiment, we noticed that a temperature shift of 1\,mK of the MZI results in an observable change of the ratio between the photon coincidence counts $N_{\rm S1-AS5}$ and  $N_{\rm S1-AS6}$. In addition, the probe process may reduce the visibility of the interference between the two terms in the superposition state, due to that detecting one S3 may project the initial state 
\begin{equation}\label{APampE0}
c_1 \left| {0} \right\rangle _{\rm{Atom}} \left| 1 \right\rangle _{{\rm{AS2}}}  + c_2 \left| 1 \right\rangle _{\rm{Atom}} \left| 0 \right\rangle _{{\rm{AS2}}}.
\end{equation}
onto a state (\ref{APampE}) with more unbalanced coefficients $c'_1$ and $c'_2$, although the probe process doesn't completely destroy or collapse the superposition state. In Fig.\,\hyperref[fig9]{9}, we present an interference only heralded by S1, where almost no detection of S3 happened. The measured data matches the fitting curves in a well way, and there is no visible deviation between them. The obtained visibility $V_2 =86\%$ in Fig.\,\hyperref[fig9]{9} is about $7\%$ higher than $V=79 \%$ obtained in Fig.\,\hyperref[fig8]{8}. We attribute the difference between these two visibilities to the detection of S3 or not, i.e. whether the wavefunction is in a projected state (\ref{APampE}) or still in the initial state (\ref{APampE0}). According to our experimental results, the detection of S3 leads to a greater imbalance in the interference. A straightforward explanation is that the detection of S3 corresponds to a lower generation probability for AS2 due to the anti-correlation between S3 and AS2, and corresponds to a higher probability for the survival (generation) of the atomic excitation heralded by S1 (S3), and thus a higher generation probability for AS4.


\begin{thebibliography}{99} 
{
\bibitem{Lundeen2011} Lundeen, J. S., Sutherland, B., Patel, A., Stewart, C. and Bamber C. Direct measurement of the quantum wavefunction. {\it Nature} {\bf 474}, 188-191 (2011).
\bibitem{Ringbauer2015} Ringbauer, M. {\it et al.} Measurements on the reality of the wavefunction. {\it Nature Phys.} {\bf 11}, 249-254 (2015). 

\bibitem{Feynman} Feynman, R., Leighton, R. and Sands, M. The Feynman Lectures on Physics Volume III, Chapter 1 (https://www.feynmanlectures.caltech.edu/III\_toc.html).

\bibitem{Scully1991} Scully, M. O., Englert, B. G. and Walther, H. Quantum optical tests of complementarity. {\it Nature} {\bf 351}, 111-116 (1991).

\bibitem{Durr1998} D\"urr, S., Nonn, T. and Rempe, G. Origin of quantum-mechanical complementarity probed by a `which-way' experiment in an atom interferometer. {\it Nature} {\bf 395}, 33-37 (1998).

\bibitem{Drossel2018} Drossel, B. and Ellis, G. Contextual Wavefunction collapse: an integrated theory of quantum measurement. {\it New J. Phys.} {\bf 20}, 113025 (2018).
\bibitem{Bild2023} Bild, M. {\it et al.} Schr\"odinger cat states of a 16-microgram mechanical oscillator. {\it Science} {\bf 380}, 274-278 (2023).

\bibitem{Bohm1952I} Bohm, D. A suggested interpretation of the quantum theory in terms of "hidden" variables. I. {\it Phys. Rev.} {\bf 85}, 166-179 (1952).

\bibitem{Oriols2019} Oriols, X. and  Mompart, J. ``Overview of Bohmian mechanics''; Chapter 1 of the book ``Applied Bohmian Mechanics: From Nanoscale Systems to Cosmology''. Editorial Pan Stanford Publishing Pte. Ltd (Second edition, 2019). 

\bibitem{Tumulka2021} Tumulka, R. Bohmian mechanics. {\it arXiv}: 1704.08017v3 (2021).

\bibitem{Sanz2019} Sanz, A. S. Bohm's approach to quantum mechanics: Alternative theory or practical picture?. {\it Front. Phys.} {\bf 14}, 11301 (2019).

\bibitem{Wiseman1998} Wiseman, H. Bohmian analysis of momentum transfer in {\it welcher Weg} measurements. {\it Phys. Rev. A} {\bf 58}, 1740-1756 (1998). 

\bibitem{Kocsis2011} Kocsis, S. {\it et al.} Observing the average trajectories of single photons in a two-slit interferometer. {\it Science} {\bf 332}, 1170-1173 (2011). 

\bibitem{Braverman2013} Braverman, B. and Simon, C. Proposal to observe the nonlocality of Bohmian trajectories with entangled photons. {\it Phys. Rev. Lett.} {\bf 110}, 060406 (2013).

\bibitem{Mahler2016} Mahler, D. H. {\it et al.} Experimental nonlocal and surreal Bohmian trajectories. {\it Sci. Adv.} {\bf  2}: e1501466 (2016).

\bibitem{Xiao2019} Xiao, Y. {\it et al.} Observing momentum disturbance in double-slit ``which-way'' measurements. {\it Sci. Adv.} {\bf  5}: eaav9547 (2019).

\bibitem{Foo2022} Foo, J., Asmodelle, E., Lund, A. P. and  Ralph, T. C. Relativistic Bohmian trajectories of photons via weak measurements. {\it Nat. Commun.} {\bf 13}, 4002 (2022).



\bibitem{Dong2020} Dong, M. X.  {\it et al.} Temporal Wheeler's delayed-choice experiment based on cold atomic quantum memory. {\it npj Quantum Information} {\bf 6}:72 (2020).

\bibitem{Zhou2017} Zhou, Z. Y. {\it et al.} Quantum twisted double-slits experiments: confirming wavefunctions' physical reality. {\it Science Bulletin} {\bf 62}, 1185-1192 (2017).

\bibitem{Wang1991} Wang, L. J., Zou, X. Y. and Mandel, L. Experimental test of the de Broglie guided-wave theory for photons. {\it Phys. Rev. Lett.} {\bf 66}, 1111-1114 (1991).
\bibitem{Croca1992} Croca, J. R. , Garuccio, A., Lepore, V. L. and Moreira, R. N. Comment on ``Experimental test of the de Broglie guided-wave theory for photons''. {\it Phys. Rev. Lett.} {\bf 68}, 3813 (1992).
\bibitem{Zou1992} Zou, X. Y., Grayson, T. Wang, L. J. and Mandel, L. Can an ``empty" de Broglie pilot wave induce coherence? {\it Phys. Rev. Lett.} {\bf 68}, 3667-3669 (1992).
\bibitem{Jeffers1994} Jeffers, S. and Sloan, J. An experiment to detect ``empty" waves. {\it Found. Phys. Lett.} {\bf 7}, 333-341 (1994).
\bibitem{Folman1995} Folman, R. and Vager, Z. Empty wave detecting experiments: A comment on auxiliary ``hidden" assumptions. {\it Found. of Phys. Lett.} {\bf 8}, 55-61 (1995).
\bibitem{Muckenheim1988} M{\"u}ckenheim, W., Lokai, P. and Burghardt, B. Empty waves do not induce stimulated emission in laser media. {\it Phys. Lett. A} {\bf 127}, 387-390 (1988).
\bibitem{Selleri1988} Selleri, F. Reply to ``empty waves do not induce stimulated emission in laser media". {\it Phys. Lett. A} {\bf 132}, 72-74 (1988).
\bibitem{Broglie1968} de Broglie, L. and Silva, J. A. E. Interpretation of a recent experiment on interference of photon beams. {\it Phys. Rev.} {\bf 172}, 1284-1285 (1968).
\bibitem{Auletta04} Auletta, G. and Tarozzi, G. On the physical reality of quantum waves. {\it Foundations of Physics} {\bf 34}, 333-341 (2004).

\bibitem{Scully1997} Scully, M. O. and Suhail Zubairy, M. Quantum optics. page 461 (Cambridge University Press, 1997).
\bibitem{Duan2001} Duan, L.-M. Lukin, M. D., Cirac, J. I. and Zoller, P. Long-distance quantum communication with atomic ensembles and linear optics. {\it Nature} {\bf 414}, 413-418 (2001).

\bibitem{Hammerer2010}  Hammerer, K., S{\o}rensen, A. S. and Polzik, E. S. Quantum interface between light and atomic ensembles. {\it Rev. Mod. Phys.} {\bf 4}, 1041-1093 (2010).

\bibitem{Louisell1961} Louisell, W. H., Yariv, A. and Siegman, A. E. Quantum fluctuations and noise in parametric processes. I. {\it Phys. Rev.} {\bf 124}, 1646-1654 (1961). 
\bibitem{Chen2009} Chen, L. Q. {\it et al.} Enhanced Raman scattering by spatially distributed atomic coherence.  {\it Appl. Phys. Lett.} {\bf 95}, 041115 (2009).

\bibitem{Aharonov2017} Aharonov, Y. {\it et al.} Finally making sense of the double-slit experiment. {\it PNAS} {\bf 114}, 6480-6485 (2017). 

\bibitem{Pfleegor1967} Pfleegor, R. L. and Mandel, L. Interference of independent photon beams. {\it Phys. Rev.} {\bf 159}, 1084-1088 (1967).
\bibitem{Paul1986} Paul, H. Interference between independent photons. {\it Rev. Mod. Phys.} {\bf 58}, 209-231 (1986).

\bibitem{Qu2021} Qu, D.-K., Wang, K.-K., Xiao, L., Zhang, X. and Xue, P. State-independent test of quantum contextuality with either single photons or coherent light. {\it npj Quantum Information} {\bf 7}, 154 (2021).
\bibitem{Griffiths2019} Griffiths, R. B. Quantummeasurements and contextuality. {\it Phil. Trans.R. Soc.} A377: 20190033 (2019).
\bibitem{Howard2014} Howard, M., Wallman, J., Veitch, V. and Emerson, J. Contextuality supplies the `magic' for quantum computation. {\it Nature} {\bf 510}, 351-355 (2014).

\bibitem{Kim2000} Kim, Y.-H., Yu, R., Kulik, S. P., Shih Y. and Scully, M. O. Delayed ``choice" quantum eraser. {\it Phys. Rev. Lett.} {\bf 84}, 1 (2000).
\bibitem{Jacques2007} Jacques, V. {\it et al.} Experimental realization of Wheeler's delayed-choice gedanken experiment. {\it Science} {\bf 315}, 966-968 (2007).
\bibitem{Peruzzo2012} Peruzzo, A., Shadbolt, P., Brunner, N., Popescu, S. and O'Brien, J. L. A quantum delayed-choice experiment. {\it Science} {\bf 338}, 634-637 (2012).
\bibitem{Kaiser2012} Kaiser, F., Coudreau, T., Milman, P., Ostrowsky, D. B. and Tanzilli, S. Entanglement-enabled delayed-choice experiment.{\it Science} {\bf 338}, 637-640 (2012).
\bibitem{Shadbolt2014} Shadbolt, P., Mathews, J. C. F., Laing, A. and O'Brien, J. L. Testing foundations of quantum mechanics with photons. {\it Nature Phys.} {\bf 10}, 278-286 (2014). 
\bibitem{Ma2016} Ma, X.-S., Kofler, J. and Zeilinger, A. Delayed-choice gedanken experiments and their realizations. {\it Rev. Mod. Phys.} {\bf 88}, 015005 (2016).
\bibitem{Chaves2018} Chaves, R., Lemos, G. B. and Pienaar, J. Causal modeling the delayed-choice experiment. {\it Phys. Rev. Lett.} {\bf 120}, 190401 (2018).
\bibitem{Wang2019} Wang, K., Xu, Q., Zhu, S.-N. and Ma, X.-S. Quantum wave-particle superposition in a delayed-choice experiment. {\it Nature Photon}. {\bf 10}, 1038 (2019). 
\bibitem{Yu2019} Yu, S. {\it et al.} Realization of a causal-modeled delayed-choice experiment using single photons. {\it Phys. Rev. A} {\bf 110}, 012115 (2019).


\bibitem{Dunningham2007} Dunningham, J. and Vedral, V. Nonlocality of a single particle. {\it Phys. Rev. Lett.} {\bf 99}, 180404 (2007).
\bibitem{Brunner2014} Brunner, N., Cavalcanti, D., Pironio, S., Scarani, V. and Wehner, S. Bell nonlocality. {\it Rev. Mod. Phys.} {\bf 86}, 419-478 (2014).
\bibitem{LiuDN2022} Liu, D.-N. {\it et al.} Generation and dynamic manipulation of frequency degenerate polarization entangled Bell states by a silicon quantum photonic circuit. {\it Chip} {\bf 1}, 1 (2022).
\bibitem{Ringbauer2016} Ringbauer, M. {\it et al.} Experimental test of nonlocal causality. {\it Sci. Adv.} {\bf 2}: e1600162 (2016).
\bibitem{Hao2022} Hao, Z. Y. {\it et al.} Demonstrating shareability of multipartite Einstein-Podolsky-Rosen steering. {\it Phys. Rev. Lett.} {\bf 128}, 120402 (2022).
\bibitem{Li2021} Li, H. {\it et al.} Heralding quantum entanglement between two room-temperature atomic ensembles. {\it Optica} {\bf 8}, 925-929 (2021).
\bibitem{Wang2022} Wang, Y. {\it et al.} Topologically protected polarization quantum entanglement on a photonic chip. {\it Chip} {\bf 1}, 3 (2022).
\bibitem{Pang2023} Pang, X.-L. {\it et al.} Entangling motional atoms and an optical loop at ambient condition. {\it npj Quantum Information} {\bf 9}, 62 (2023).

\bibitem{Pan2023} Pan, Y.-M., Cohen, E., Karimi, E. {\it et al.} Weak measurements and quantum-to-classical transitions in free electron-photon interactions {\it Light: Science $\&$ Applications} {\bf 12}:267 (2023).
\bibitem{Pan2020} Pan, Y.-M., Zhang, J., Cohen, E. {\it et al.} Weak-to-strong transition of quantum measurement in a trapped-ion system. {\it Nat. Phys.} {\bf 16}, 1206-1210 (2020).

\bibitem{Lijiakun2023} Li, J.-K., Sun, K., Wang, Y. {\it et al.} Experimental demonstration of separating the wave-particle duality of a single photon with the quantum Cheshire cat. {\it Light: Science $\&$ Applications} {\bf 12}:267 (2023).


\bibitem{Nunn2008} Nunn, J. Quantum memory in atomic ensembles. Ph.D. thesis, University of Oxford, 481-487 (2008).
\bibitem{Dou2018} Dou, J. P. {\it et al.} A broadband DLCZ quantum memory in room-temperature atoms. {\it Commun. Phys.} {\bf 1}, 55 (2018).

\bibitem{Liu2022} Liu, C., Ye, H.-F. and Shi, Y.-L. Advances in near-infrared avalanche diode single-photon detectors.  {\it Chip} {\bf 1}, 5 (2022).
\bibitem{Hu2022} Hu, A.-Q., Liu, Q.-L. and Guo, X. Carrier localization enhanced high responsivity in graphene/semiconductor photodetectors. {\it Chip} {\bf 1}, 6 (2022).



}
\end{thebibliography}
\end{document}